\documentclass[elsart12]{elsart}
\input{psfig}
\usepackage{amssymb}

\newcommand{\cy}{{\mathcal Y}}
\newcommand{\cp}{{\mathcal P}}

\begin{document}
\begin{frontmatter}

\title{Parameter Estimation of Hidden Diffusion Processes:
Particle Filter vs. Modified Baum-Welch Algorithm}
\author{A. Benabdallah$^{(a,b)}$ and G. Radons$^{(b)}$}
\address{$^{(a)}$ Max-Planck-Institut f\"ur Physik komplexer Systeme,
N\"othnitzer Stra\ss e 38, D-01187 Dresden, Germany \\}
\address{$^{(b)}$Institute of Physics, Chemnitz University of
Technology, D-09107 Chemnitz, Germany}

\date{\today}

\begin{abstract}
We propose a new method for the estimation of parameters of 
hidden diffusion processes. 
Based on parametrization of the transition matrix, 
the Baum-Welch algorithm is improved. 
The algorithm is compared to the particle filter
in application to the noisy periodic systems.
It is shown that the modified Baum-Welch algorithm
is capable of estimating the system parameters with better
accuracy than particle filters.
\end{abstract}


\end{frontmatter}


\def\figsize{3.5in}

\maketitle

\section{Introduction}
Identification or parameter estimation is one of the most important
and interesting fields in the nonlinear dynamics and time series
analysis. There exist many methods of identifying the parameters of
a nonlinear stochastic system such as maximum Likelihood estimators
and Bayes estimators \cite{Casella:2001}.
Especially this latter is related to the Sequential Monte Carlo methods
\cite{Kitagawa:1996} which are also known as Particle filter methods
introduced by  Gordon et al.  \cite{Gordon:1993}.
These methods utilize a large number of random samples (or particles)
to represent the posterior probability distributions. The particles are
propagated over time using a combination of sequential importance
sampling and resampling steps. At each time step the resampling procedure 
statistically multiplies and/or discards particles to adaptively concentrate
particles  in regions of high posterior probability.\\
Particle filter methods are usually applied to state space models 
to approximate the posterior probability distributions of the state 
given the available observation.
If the state space models contain a set of unknown parameters which 
are to be estimated then one can include them in the model by augmenting 
the state vector \cite{Liu:2001}.\\
As an alternative method for estimating parameters of continuous hidden
diffusion processes we propose 
a hidden Markov models (HMMs) \cite{Rabiner:1989} approach which model
both the signal and noise simultaneously \cite{Elliott:1995}.
This is based on the approximation of continuous systems by 
discrete models.
The underlying signal is assumed to be generated by a discrete 
Markov chain. The latter uses the joint probability of the sequence
of the discrete observation samples as the likelihood function.
The general theory of HMMs was established by Baum et al. 
in the sixties \cite{Baum:1966,Baum:1967,Baum:1970}.\\ 
Standard HMMs rely on the Baum-Welch reestimation procedure to optimize the
likelihood function \cite{Rabiner:1989}. 
The standard Baum-Welch algorithm suffers from the problem that it may
converge to a local minimum. However, we can overcome this
difficulty by parameterizing the transition matrix \cite{Ben:2004}.\\
Previous works have shown that hidden Markov models are successful tools for modeling
and classifying dynamic behaviors. For example, HMMs are used for analyzing 
biological sequences \cite{Durbin:2001}, speech recognition \cite{Rabiner:1993},
ion channel analysis \cite{Shin:1990,Fredkin:1992,Becker:1994}, 
and to detect different modes of neuronal activity \cite{Radons:1994}.\\
In experimental physics, the objective of any measurement is to 
determine the value of the particular quantity to be measured.
In general, however, the result of a measurement is only an
approximation or estimate of the value one is looking for.
For instance in coupled Josephson junctions \cite{Likharev:1986}, 
 direct measurements of the time dependence of the voltage are 
usually impossible because the characteristic time scale of 
voltage variations is too short ($\sim$ picoseconds).
One can usually measure the Josephson radiation emission in some 
narrow frequency range, which can show chaotic
behaviour. But in this case one cannot
see higher harmonics which are required to fully reconstruct the
voltage time evolution. In experiments, another version
of the voltage is usually measured which is the results of the 
low pass filtering.
The obtained voltage is used to extract the current-voltage characteristic. 
In this case, the observed variable is the voltage whereas the 
Josephson phase is hidden.
Likharev \cite{Likharev:1986} reported that the coupled Josephson junctions 
belong to a class of complex systems. The corresponding experimental works 
show that it is difficult to estimate some of the parameters characterizing the Josephson device, e.g.
the damping related to the fluctuation of the temperature and the
maximal Josephson current.\\  
For the case that the measured time series proved to be approximately Markovian, 
Friedrich et al \cite{Siegert:1998,Friedrich:2002} proposed an 
approach to obtain the drift and diffusion of one-dimensional 
Langevin equations from the time series. This is based on the 
finite-difference form of their definition together with suitable
interpolations of the resulting trends. Ragwitz et al. \cite{Ragwitz:2001} 
proposed a correction of this approach to reduce the errors 
due to a finite times step. 
This was controversy for the case of 
directly observed states of continuous diffusion
processes, which were measured in discrete time \cite{Ragwitz:2001},
\cite{Friedrich:2002}, \cite{Ragwitz:2002}. 
We believe that our approach can then be used to clarify this situation.\\     
\noindent
This paper is devoted to a numerical evaluation of these two methods 
by applying them, for instance, to the problem of diffusion in periodic 
potentials 
with noisy observations. The latter example is taken as a periodic function of the 
coordinate
of the diffusing state. The aim of this work is to estimate the drift coefficient 
and the diffusion  constant.\\
The paper is organized as follows. In the next section, we formulate the
problem of hidden diffusion processes. In section 3, we review the particle filter
and propose a modified Baum-Welch algorithm. Section 4 is devoted to a numerical 
simulation and the evaluation of both methods. 
Conclusions are given in the last section. 

\section{Mathematical model} 
Diffusion processes are usually modeled by the evolution equation of the 
probability density function $\cp$ which is governed by the continuous 
Fokker-Planck equation. This can be read in one dimension 
\begin{equation}
\label{fke}
\frac{\partial} {\partial t} {\cp}(x,t)  =  - \frac{\partial } 
{\partial x} \left[ F(x) {\cp}(x,t) \right] 
+ {1 \over 2} \frac{\partial^2} {\partial x^2} \left[ D(x) {\cp}(x,t)\right]
\;, 
\end{equation}
where $F$ is the drift and $D$ is the diffusion coefficient.
\noindent
The process of Eq. (\ref{fke}) can equivalently be described by a Langevin equation
interpreted in It\^{o} sense
\begin{equation}
\label{langeq}
\dot x = F(x) + \sqrt{D(x)}\, v(t) \;, 
\end{equation}
where $v(t)$ is intrinsic white noise with the density $q(v_t)$, 
and the initial condition of Eq. (\ref{langeq}) is given by 
$x(t=0) ~\sim \mu(x(0))$.\\ 
\noindent
The state variable $x(t)$ can usually not be observed directly but only via 
a measurement process, which is modeled by an observation function $h$ as  
\begin{equation} 
\label{obse}
y(t) = h(x(t),w(t)) \; ,
\end{equation}
where $w(t)$ denotes observation noise with the density $r(w_t)$ which is 
independent of $v(t)$.\\ 
Eq. (\ref{langeq}) together with Eq. (\ref{obse}) define  
hidden diffusion processes.\\ 
In general, the observation function $h$ is nonlinear. 
Thus, the diffusing state $x$ is {\it hidden}. The estimation problem therefore 
becomes difficult to tackle and 
there exists no analytical method dealing with diffusion 
coefficient estimation.
There exist only approximate numerical methods such as particle filters.
\noindent
In practice, the particle filters are applied to discretized version of the system 
(\ref{langeq}--\ref{obse}), which result in the state space model
\begin{eqnarray}
\label{st_eq}
x_t &=& G_t(x_{t-1}, v_t)  \\ \label{ob_eq}
y_t &=& H_t(x_t,w_t)\, , 
\end{eqnarray}
where $G_t$ and $H_t$ are assumed to be known nonlinear functions, the dynamical white noise 
$v_t$ and the observation white noise $w_t$ are independent random processes and
the initial condition $x_0 \sim p(x_0)$. \\
In the next section, we give a short overview of the particle filters and 
the corresponding  implementation issues (more details can be found in 
\cite{Gordon:1993,Kitagawa:1996,Arulampalam:2002,Doucet:2001}).\\ 
We assume that the diffusion coefficient $D(x)=D$ is constant
throughout the paper.\\
\section{Algorithms}
\subsection{ Particle filter algorithm (Monte Carlo Filter)}
Consider systems that are described by the generic state space model 
(\ref{st_eq}--\ref{ob_eq}). Sequential Monte Carlo methods or particle filters
provide an approximate 
Bayesian solution to the discrete time recursive 
of the state space model (\ref{st_eq}--\ref{ob_eq}) by updating an approximate 
description of the posterior filtering
density.\\
\noindent
Let $x_t$ denote the state of the observed system and 
$\cy_t=\{y_i\}_{i=1}^{t}$ the set of observations up to the present 
time $t$. Let the independent process noise $v_t$ and the measurement noise $w_t$
with the densities $q(v_t)$ respective $r(w_t)$. The initial uncertainty is
described by the density $p(x_0)$.
The particle filter approximates the probability density  $p(x_t|\cy_t)$
by using a large set of $N_p$ particles $\{x^{(i)}_t\}_{i=1}^{N_p}$, where
each particle has an assigned relative weight $m^{(i)}_t$, such that all
weights sum to one. The particle filter updates the particle location 
and the corresponding weights recursively with each new observation. The 
non-linear prediction density  $p(x_t|\cy_{t-1})$ and optimal filtering 
density  $p(x_t|\cy_{t})$ for the Bayesian interference are given by
\begin{eqnarray}
\label{pdf1}
p(x_t|\cy_{t-1}) & = &  \int p(x_t|x_{t-1})p(x_{t-1}|\cy_{t-1}) {\rm d} x_{t-1} \\
\label{pdf2}
p(x_t|\cy_{t}) &  = & \frac{p(y_t|x_t)\,p(x_t|\cy_{t-1})}{p(y_t|\cy_{t-1})}\;,
\end{eqnarray}
where $p(y_t|\cy_{t-1})=\int p(y_t|x_t) p(x_t|\cy_{t-1}) {\rm d} x_t$. 
The transition probability density $p(x_t|x_{t-1})$ is know as the motion model
(\ref{st_eq}) and $p(x_{t-1}|\cy_{t-1})$ is the updated estimate from the 
previous step. $p(y_t|x_t)$ is the observation probability density given by  
Eq. (\ref{ob_eq}).\\ 
Note that, generally, these equations are not analytically tractable. However, 
for the important special case of linear dynamics, linear measurements and 
Gaussian noise there exist a closed form solution of Eq.(\ref{pdf1}--\ref{pdf2}) , given by the Kalman filter \cite{Kalman:1960}.\\ 
\noindent
The main idea of the optimal filter is to approximate $p(x_t|\cy_{t-1})$
with 
\begin{equation}
\label{pdf3}
p(x_t|\cy_{t-1}) \approx {1 \over N_p}  \sum_{i=1}^{N_p} \delta(x_t - x^{(i)}_t)
\,,
\end{equation}
where $\delta$ is the Dirac delta distribution.\\ 
Inserting (\ref{pdf3}) into (\ref{pdf2}) yields a density of a simple form. This 
can be done by using
the Bayesian bootstrap or Sampling Importance Resampling (SIR) algorithm from 
\cite{Kitagawa:1996} which is given by the following algorithm 

\noindent
{\framebox{\parbox[c]{14.5cm}{
\begin{enumerate}
\item At $t=0$, generate random numbers $x_0^{(j)} \sim p(x_0) \, 
{\sf for} \, j=1,...,N_p$ 
\item Repeat the following steps for $t=1,...,T$ 
\begin{enumerate}
\item Generate random numbers $v_t^{(j)} \sim q(v) \, {\sf for} \, j=1,...,N_p$ 
\item Compute $p_t^{(j)} = G(x_{t-1}^{(j)}, v_t^{(j)}) \, {\sf for} \, j=1,...,N_p$  
\item Compute $m_t^{(j)} = p(y_t|p_t^{(j)}) \, {\sf for} \, j=1,...,N_p$ 
\item Resample with replacement $N_p$ particles $\{x_t^{(j)}\}$ 
from $\{p_t^{(j)}\}$ according to the importance weights
\end{enumerate}
\end{enumerate}
}}
\\

\centerline {Table 1: Particle filter algorithm}
}
\vspace{0.5cm}
\noindent 
Note that the resampling procedure (step (2d) in the Table 1) selects only the fittest 
particles to obtain an unweighted measure $\{(x_t^{(j)},\frac 1 {N_p}\}$.\\ 
\noindent
Sometimes the resampling step is omitted and simply imposed when needed to
avoid a divergence in the particle filter as in the sequential importance
sampling (SIS) method, where the weight is updated recursively as \cite{Doucet:2001}
\[
m_t^{(j)} = m_{t-1}^{(j)} \cdot p(y_t|p_t^{(j)})\, {\sf for} \, j=1,...,N_p
\]
As the estimate of the state we choose the minimum mean square estimate, i.e. 
\begin{equation}
\hat x_t  =  \int x_t p(x_t|\cy_{t}) {\rm d} x_t \approx \sum_{i=1}^{N_p} 
\tilde{m}^{(i)}_t x^{(i)}_t \, . \\
\end{equation}
where $\tilde{m}^{(i)}_t = {m}^{(i)}_t/\sum_{i=1}^{N_p} {m}^{(i)}_t$.\\

\noindent
{\it Parameter estimation}:
the state space model (\ref{st_eq}--\ref{ob_eq}) usually contains several unknown
parameters, such as the variances of the noises and the coefficients of 
the functions $F_t$ and $H_t$. Let us denote such unknown parameters by 
${\theta} =(\theta_0,\cdots,\theta_{N_\theta})$. We consider a Bayesian 
estimation problem by augmenting the state vector $x_t$ with the unknown parameter 
vector ${\theta}$ as
\begin{equation}
z_t = 
\left [
\begin{array}{l}
x_t \\
{\theta_t}
\end{array}
\right ] \, ,
\end{equation}
with $\dot {{\theta}_t} = 0 $. 
The state space model for this augmented state vector $z_t$ is thus
\begin{equation}
\label{ssma}
\begin{array}{ll}
z_t = G^*_t(z_{t-1},v_t) \\
y_t = H^*_t(z_t,w_t)\quad ,
\end{array}
\end{equation}
where the nonlinear functions $G^*_t(z,v) = (G_t(x,v), \theta_t)$ and 
$H^*_t(z,w) = H_t(x,w)$. We can therefore apply the particle
filter algorithm to the state space models described by Eq. (\ref{ssma}) 
as previously.
\subsection {HMM and Modified Baum-Welch algorithm}
The approximation of continuous hidden diffusion processes 
(\ref{langeq})--(\ref{obse}) by discrete models results in Hidden Markov Models
(HMMs). In \cite{Ben:2004} the diffusion process 
was approximated by a discrete random walk with $N$ states with
only nearest neighbor transitions.
For the observation process we considered an appropriate discrete
process, which is well defined in the continuous time limit, 
e.g. consider
Eq. (\ref{obse}) in discrete form.\\ 
Comparing Fokker-Planck equations on the one hand with discrete time and space 
master equations on the
other hand, it is easy to establish the connection between the 
continuous diffusion process and the Markov model parameters as in \cite{Ben:2004}
\begin{eqnarray}
\label{fa}
F(i\,\Delta x)&=&{{a_{i,i+1} - a_{i,i-1}} \over \Delta x} D_0 \\ \nonumber
D(i\,\Delta x)&=&[ (a_{i,i+1} + a_{i,i-1}) - (a_{i,i+1} -a_{i,i-1})^2] D_0 \, , \\ \label{da}
\end{eqnarray}
where $D_0= {{\Delta x}^2 \over {\Delta t}}$ and $a_{i,j}$ are the elements
of the transition matrix. The continuum limit 
${\Delta x},{\Delta t} \rightarrow 0$ can be approached by keeping $D_0$ constant.\\
Relations (\ref{fa}--\ref{da}) are important, because they give a justification for
the approximation of continuous diffusion processes by discrete models.\\ 
Standard HMMs rely on standard Baum-Welch reestimation procedure to 
optimize the 
likelihood function (more details can be found in \cite{Ben:2004}).
The  procedure may have several drawbacks 
if it is applied to the problem of diffusion in periodic potentials 
with noisy observations.  Since the observation
function was simply chosen as the cosine of the state variable,
the maxima of likelihood function are degenerate e.g. each state is observed 
with two different observations.
More importantly, in order to converge to the continuous
hidden diffusion processes, we should choose a large number of states,
which means that many parameters should be re-estimated.
In this case, the standard Baum-Welch algorithm is not
applicable, due to the limited number of observations.
To avoid these problems we have to use a modified version of the Baum-Welch algorithm . 
It consists in parameterizing the matrix of the transition probability. For instance,
this can be done by a Fourier expansion of the elements of the transition matrix.
\begin{equation}
\label{star}
\{a_{k,k+j}\} = \{a_{k,k+j}({\theta})\} = \sum_{n=0}^{N-1} 
\theta_n(j) exp\left(i \,2\pi \, \frac {kn}{N} \right) \qquad {\rm for}\, j=-1,1\; ,
\end{equation} 
Following \cite{Michalek:1999} we obtain $N_\theta$ nonlinear implicit equations
\begin{equation}
\label{jens}
\sum_{i \neq j} \frac{\partial a_{ij}({\theta}^{(n+1)})}
{\partial \theta_\kappa^{(n+1)}} \left (
\frac{\Psi_{ij}({\cy}_t, \{a_{ij}({\theta}^{(n)})\})}{a_{ij}
({\theta}^{(n+1)})} 
- \frac{\Psi_{ii}({\cy}_t, \{a_{ii}
( {\theta}^{(n)})\})}{a_{ii}
({\theta}^{(n+1)})}\right ) = 0\;,
\end{equation}
for $\kappa = 1,\cdots, N_\theta$.\\
\noindent
The calculation of the conditional probability 
$\Psi_{ij}({\cy}_t, \{a_{ij}({\theta}^{(n)})\})$ at iteration $n$ 
can be carried out by using the forward-backward algorithm given in \cite{Rabiner:1989}.\\
\noindent
In case of homogeneous random walks
we derived in \cite{Ben:2004} an explicit expression for the new estimates of the
parameterized transition probabilities in terms of previous estimates
and the observed signal. In general, 
Eq. (\ref{jens}) has to be solved numerically, for example, by 
using the Newton methods. Then one can find the fixed point
${\theta}^*$ solution of Eq. (\ref{jens}).\\
Given a set of observation data ${\cy}_t=\{y\}_{t=1}^{t}$, the core of 
the modified Baum-Welch algorithm reads \\ 

\noindent
{\framebox{\parbox[c]{14.5cm}{
\begin{enumerate}
\item Generate a hidden Markov model ``trainer'' with $N$ states and 
$M$ distinct 
observation symbols   
\begin{enumerate}
\item Generate a tridiagonal transition matrix according to Eq. (\ref{star})   
\item The observation matrix is given by the time discretization of Eq. (\ref{obse})  
\end{enumerate}
\item Repeat until 
$\left | \log P(\vec y | a_{ij}({\theta}^{(n)})
-\log P(\vec y | a_{ij}({\theta}^{(n-1)}) \right | \leq 10^{-6}$ 

\begin{enumerate}
\item Compute the conditional probability $\Psi_{ij}({\cy}_t, \{a_{ij}({\theta}^{(n)})\})$
using the forward-backward algorithm given in \cite{Rabiner:1989} 
\item Update the elements of the transition matrix using the formula (\ref{jens}) 
\end{enumerate}
\end{enumerate}
}} 
\\

\centerline {Table 2: Combined HMM and modified Baum-Welch algorithm}
}
\noindent
An obvious advantage of the modified Baum-Welch algorithm 
is that it is independent of the number of states $N$ but depends only on  
the number of Fourier coefficients. 

\section{Simulation Results}
We now present a simple example to illustrate the central ideas in this paper.
We consider the system
\begin{eqnarray}
\label{stp}
\dot X_t &=& \left(\theta_0 + \sum_{n=1}^{N_\theta} \theta_n \sin(2n \pi X_t/L)
\right) \, 
+ \sqrt{D}\,  v_t \\ \label{obp}
Y_t &=& (\cos(2 \pi X_t/L))  + \sqrt{\sigma}\, w_t \, ,
\end{eqnarray}
with the initial condition $X_0 \sim {\mathcal N}(0,1)$. The driving
noise $v_t$ and the observation noise $w_t$ are independent Gaussian random
processes of variance one. In Eq.(\ref{stp}--\ref{obp}), $L$ is the spatial extension 
(period) equal to $N \Delta x$ ( $N$ is the number of states of the discrete model) and
${\theta} = (\theta_0,\cdots,\theta_{N_\theta}, D)$ is the set of parameters 
to be estimated. 
Eq. (\ref{obp}) describes the observation processes.\\
For a practical  implementation of the particle filter and the 
modified Baum-Welch algorithm, the necessary sample paths and stochastic 
integrals must be discretely approximated. Appropriate numerical methods
are discussed by Kl\"oden and Platen \cite{kloeden:1999}. 
The Euler scheme is used here for this aim.\\
Once the observation sequence is generated by the model 
(\ref{stp}--\ref{obp}), we apply the two algorithms to reestimate the 
drift term and the diffusion constant.\\
\noindent
Note that the application of particle filters in 
estimating parameters requires regarding the set of parameters 
${\theta}$ as time dependent. That is, we have to consider
a different model in which ${\theta}$ is replaced by ${\theta}_t$ at
time $t$, and to include ${\theta}_t$ in the augmented state
vector. Then we add an independent, zero-mean normal increment to the parameters 
at each time step. As a result, 
the discretized equations of system (\ref{stp}--\ref{obp}) read 
\begin{equation}
\label{aug}
\begin{array}{llll}
x^t&= x^{t-1}+ \left(\theta_0^{t-1} + \sum_{n=1}^{N_\theta} 
\theta_n^{t-1} \sin(2n \pi X_t/L)\right) \Delta t 
+ \theta_{N_\theta +1}^{t-1} v_t\\
\theta_n^t  &=  \theta_n^{t-1} + u^t_n  \quad {\rm for} \quad 
n=0,1,\cdots, N_\theta +1 \\
y_t  &= \cos(2\pi x^t/L) + \sqrt{\sigma} \,w_t \,,\qquad \quad \qquad \qquad 
\end{array}
\end{equation}
where ${\left(\theta_{N_\theta +1}^t\right)}^2 = D_t$, 
$v_t \sim {\mathcal N} (0,\sqrt{\Delta t})$,  
$u^t_n \sim {\mathcal N} (0,\epsilon \sqrt{\Delta t})$ and 
$w_t \sim {\mathcal N} (0,1)$.\\

\noindent
For simplicity, we restrict ourselves to the case $N_\theta =1$ 
and $\Delta t=1$.\\ 
In order to compare the numerical results given by the particle
filter with the modified Baum-Welch algorithm we consider the
drift parameters $\theta_0=-0.1$, $\theta_1=0.1$, the 
diffusion constant $D=0.8$ and we assume that there is no 
observation noise $\sigma = 0$.\\ 
First, the particle filter is applied to the entire augmented state vector, 
using the scheme of Table 1. 
The initial value and the initial covariance of the estimated augmented 
state vector (\ref{aug}) we were set to

\begin{equation}
\left (
\begin{array}{cc}
\hat x^0 \\
\hat \theta^0_0 \\
\hat \theta^0_1 \\
\hat \theta^0_2 \\
\end{array}
\right ) =
\left (
\begin{array}{cc}
1 \\
0 \\
0 \\
0 \\
\end{array}
\right ) , \qquad
P^0 = \left (
\begin{array}{cccc}
5^2 & 0 & 0 & 0\\
0   & .1^2 & 0 & 0\\
0   & 0 & .1^2 & 0\\
0   & 0 & 0 & .1^2\\
\end{array}
\right )\; .
\end{equation}
The actual initial value of the state vector was drawn randomly from 
${\mathcal N}(0,P^0)$.
\begin{figure}[bt]
\centerline{\psfig{figure=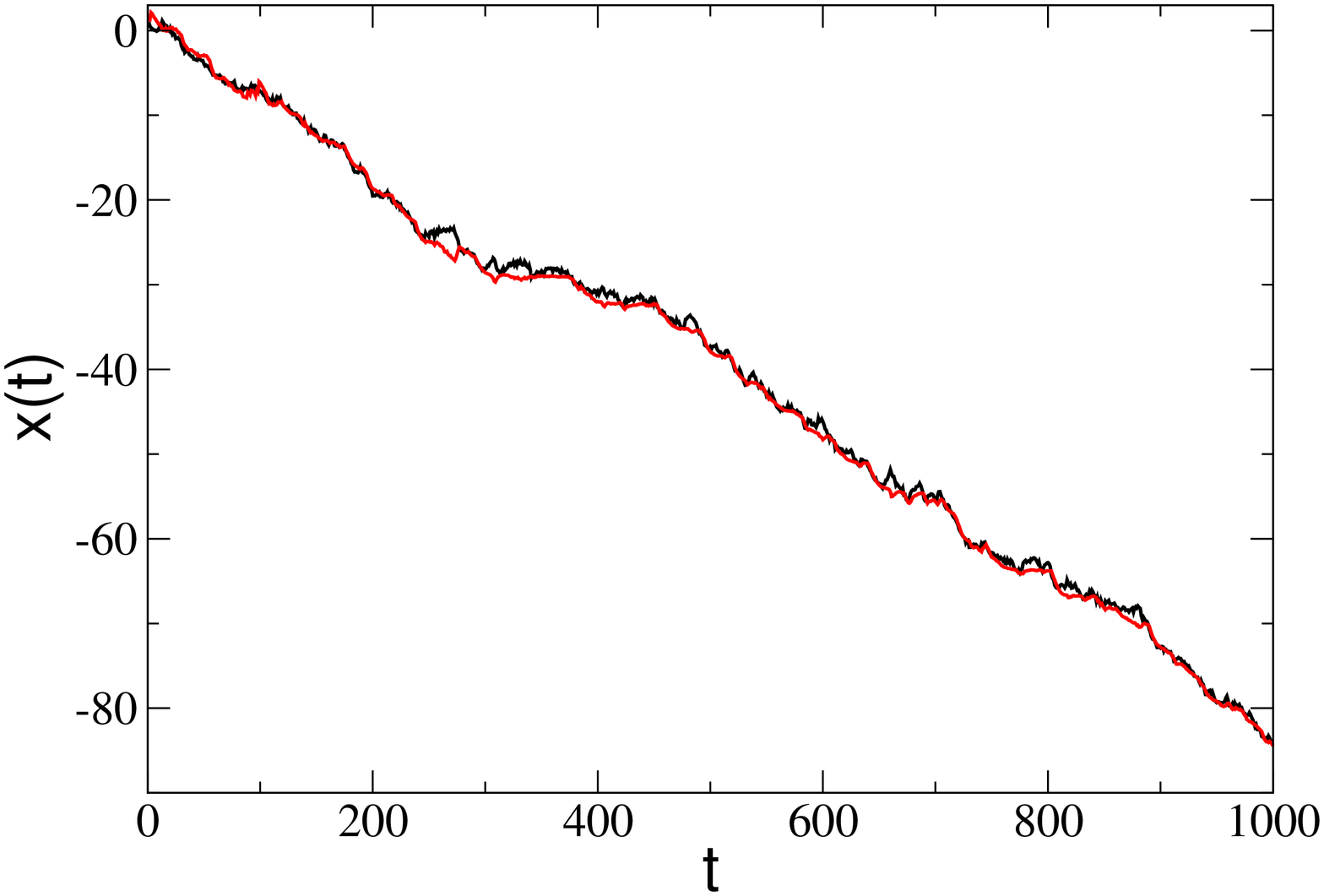,angle=-90,width=7cm,height=6cm}
\psfig{figure=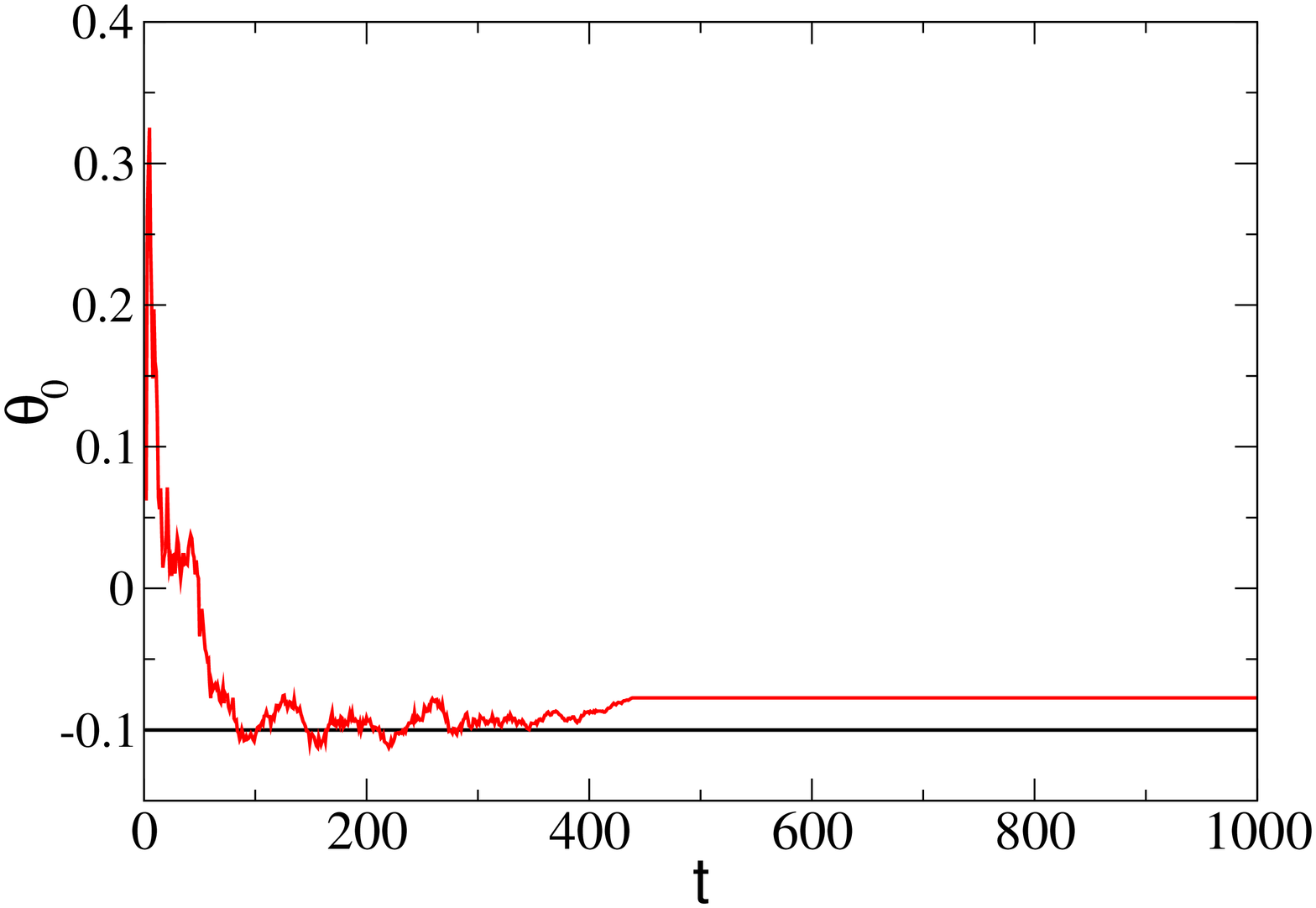,width=7cm,height=6cm,angle=-90}}
\centerline{\psfig{figure=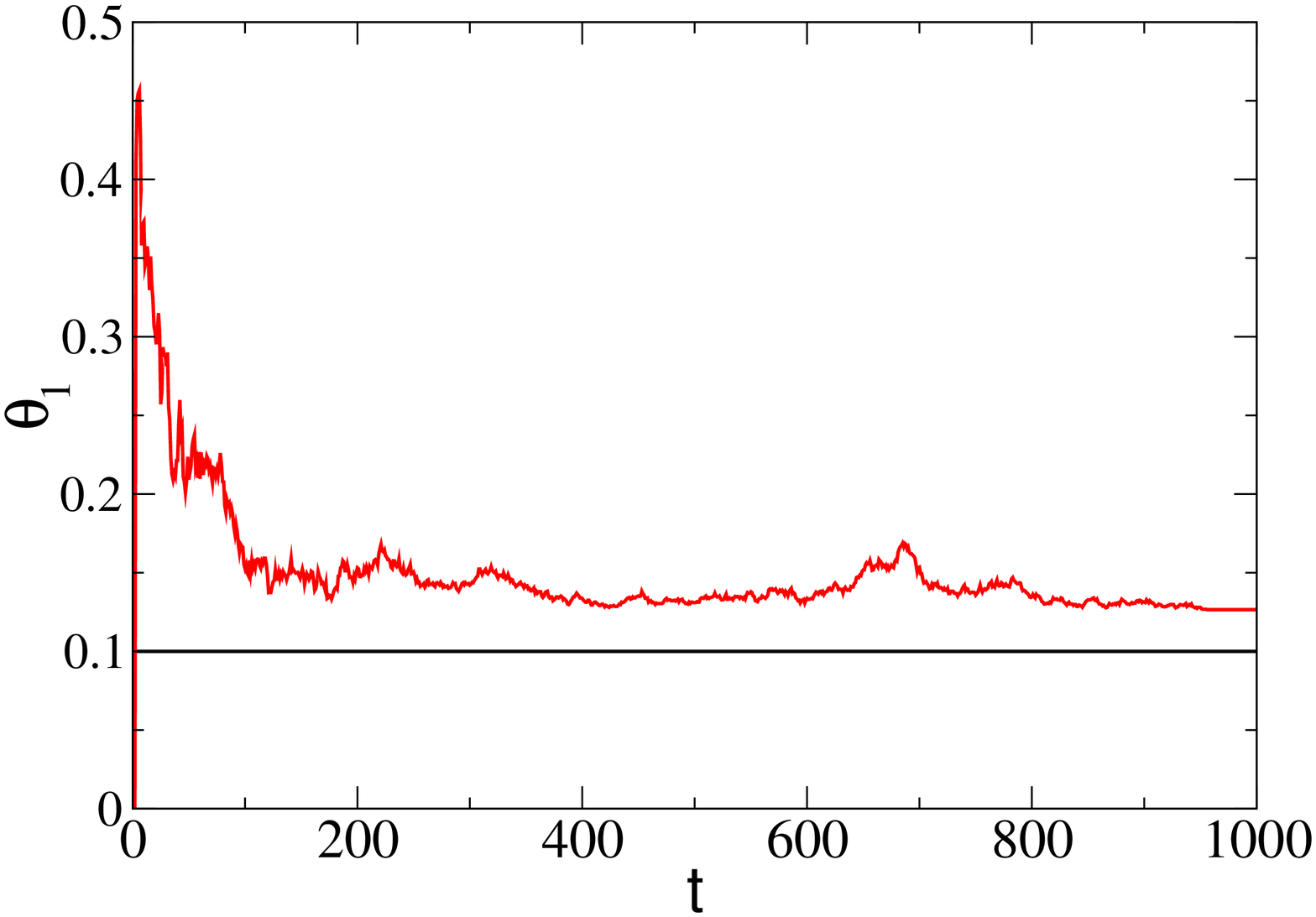,width=7cm,height=6cm,angle=-90}
\psfig{figure=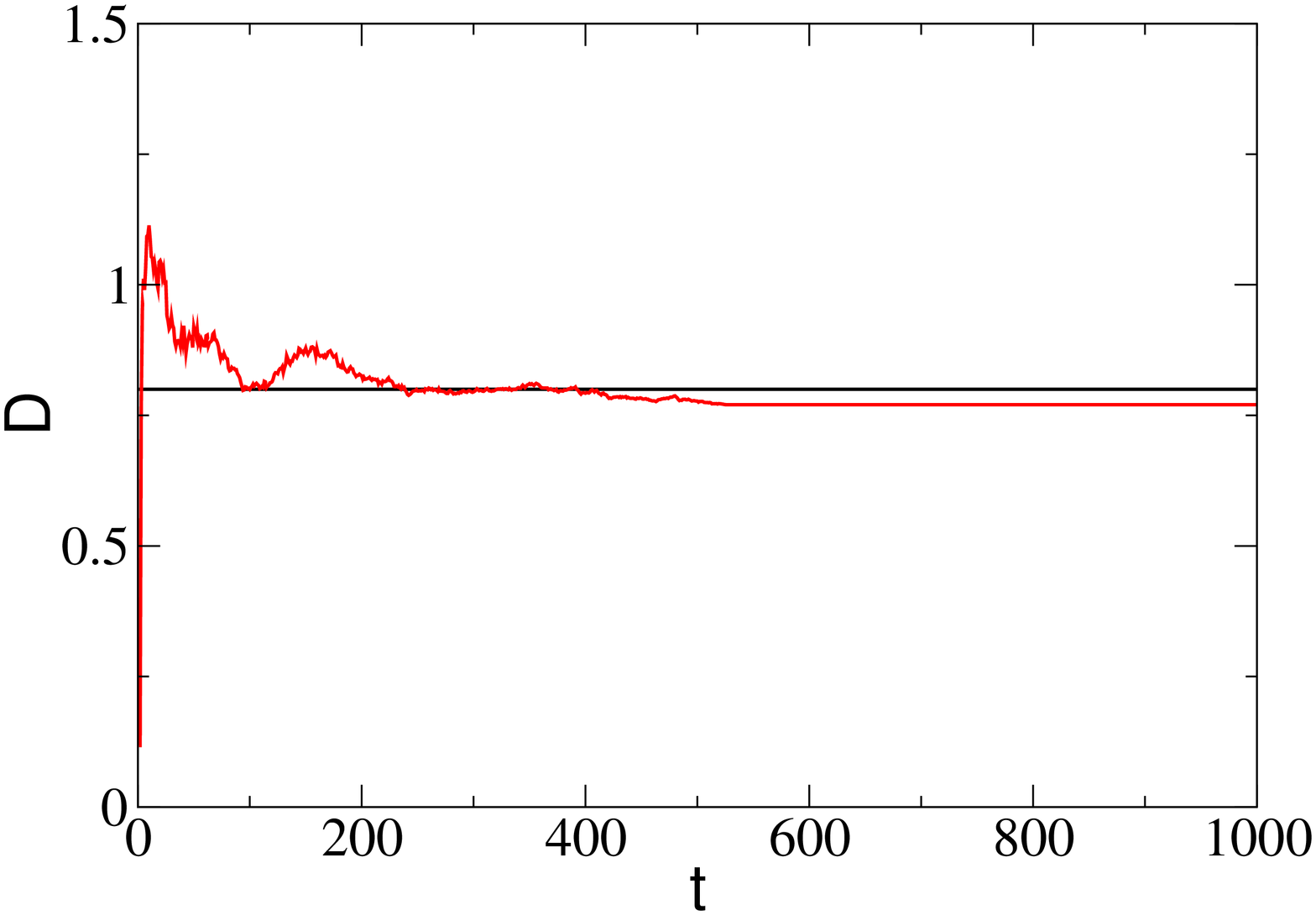,width=7cm,height=6cm,angle=-90}}
\caption{State $x$ (top left panel), parameter $\theta_0$
(top right panel), parameter $\theta_1$ (bottom left panel) and
diffusion coefficient $D$  (bottom right panel) vs. time. The correct
values are shown by solid black lines and the estimated values after applying
the particle filter algorithm are represented by solid red lines.}
\end{figure}
\noindent
Fig.~1 shows the true state $x$, the parameters $\theta_0$, $\theta_1$ and
the diffusion coefficient $D$ as a function of time and 
represent it as black solid lines. 
The values estimated from $N_p=1000$ are shown by red solid lines. After convergence, 
the particle filter
gives a  ``reasonable'' estimation for the state and better estimate 
of the correct values of the drift parameters 
$(\theta_0,\theta_1)$ and the diffusion constant $D$. However, 
the estimate state $x$ does not totally agree with the true values.\\
Note in Fig.~1 the stochastic character of the particle filter (because 
it is based on Monte Carlo methods).
\begin{figure}[bt]
\centerline{\psfig{figure=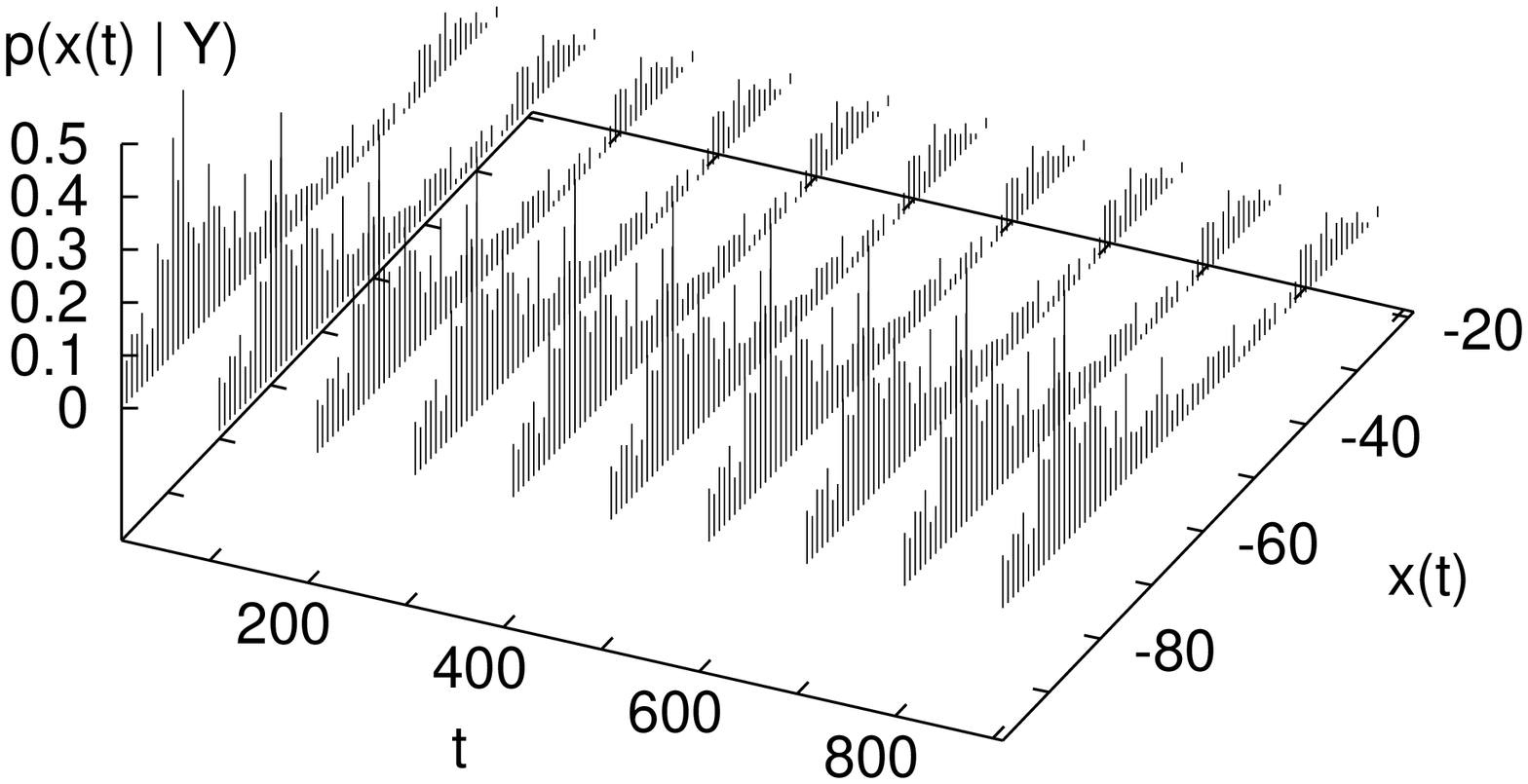,width=12cm,height=10cm,angle=-90}} 
\caption{Probability density function given by the particle filter algorithm}
\end{figure}
\noindent
Fig.~2 presents the estimated filtering distributions. One can clearly see 
from this figure the multimodal non-Gaussian posterior distribution character.\\
Moreover, Tables 3 and 4 show the performance of the 
particle filters as function of the number of particles
for two lengths of observation, $T=100$ and $T=1000$. 
More specifically, each table shows how many runs out of 
a total of 100 simulations diverged.\\ 
\begin{table}
\centerline{T=100}
\vspace{0.1cm}

\begin{center}
\begin{tabular}{|c|l|l|l|}
\hline
Number of particles & 100 &500 & 1000 \\ \hline
$\theta_0$          & 92\% & 73\% & 43\% \\ \hline
$\theta_1$          & 88\% & 74\% & 45 \%\\ \hline
$D$                 & 91\% & 76\%& 48 \%\\ \hline
\end{tabular}\\
\vspace{0.2cm}

Table 3: Percentage of diverged runs of the estimated parameters
for the particle filter.
\end{center}
\end{table}

\begin{table}
\centerline{T=1000}
\vspace{0.1cm}

\begin{center}
\begin{tabular}{|c|l|l|l|}
\hline
Number of particles & 100 &500 & 1000 \\ \hline
$\theta_0$          & 89\% & 25\% & 5\% \\ \hline
$\theta_1$          & 86\% & 25\% & 6\% \\ \hline
$D$                 & 90\% & 69\%& 10\%\\ \hline
\end{tabular}\\
\vspace{0.2cm}

Table 4: Percentage of diverged runs of the estimated parameters
for the particle filter.
\end{center}
\end{table}
One clearly sees from tables that it takes many particles and a large 
number of iterations for the particle filter to work well. The main reason 
for this is well known the degeneracy of
particle filter if the process noise has a small variance 
\cite{Doucet:2001}.\\ 

\noindent
In order to use a discrete HMM, we must first quantize the 
observation data into a set of standard vectors according to Elliott 
\cite{Elliott:1995}. 
The quantized data are used as training sets for a HMM which has to learn the
correct parameters from these observations.\\ 
Here, we ~have ~implemented ~the ~modified Baum-Welch algorithm 
described in {Table 2}. More details on implementation 
issues can be found in \cite{Ben:2004}.\\
 
Fig.~3 shows the drift function and diffusion 
constant as a function of the coordinate $x$. The estimated values 
are represented as dot-dashed lines after applying 
the particle filter and as dashed lines for  the modified Baum-Welch 
algorithm.\\  
One can see from this figure that a convergence of the modified Baum-Welch algorithm 
to the correct parameter values was obtained.
Moreover, the convergence is very fast, $n \approx 15$ ($n$ is the number of iterations),
whereas the particle filter algorithm
needs a large number of particles, $N_p=1000$, and needs larger number of iterations 
(around $500$) until convergence is obtained. Therefore, the time consuming is 
more relevant for the particle 
filter algorithm than for the modified Baum-Welch algorithm.\\
\begin{figure}
\centerline{\psfig{figure=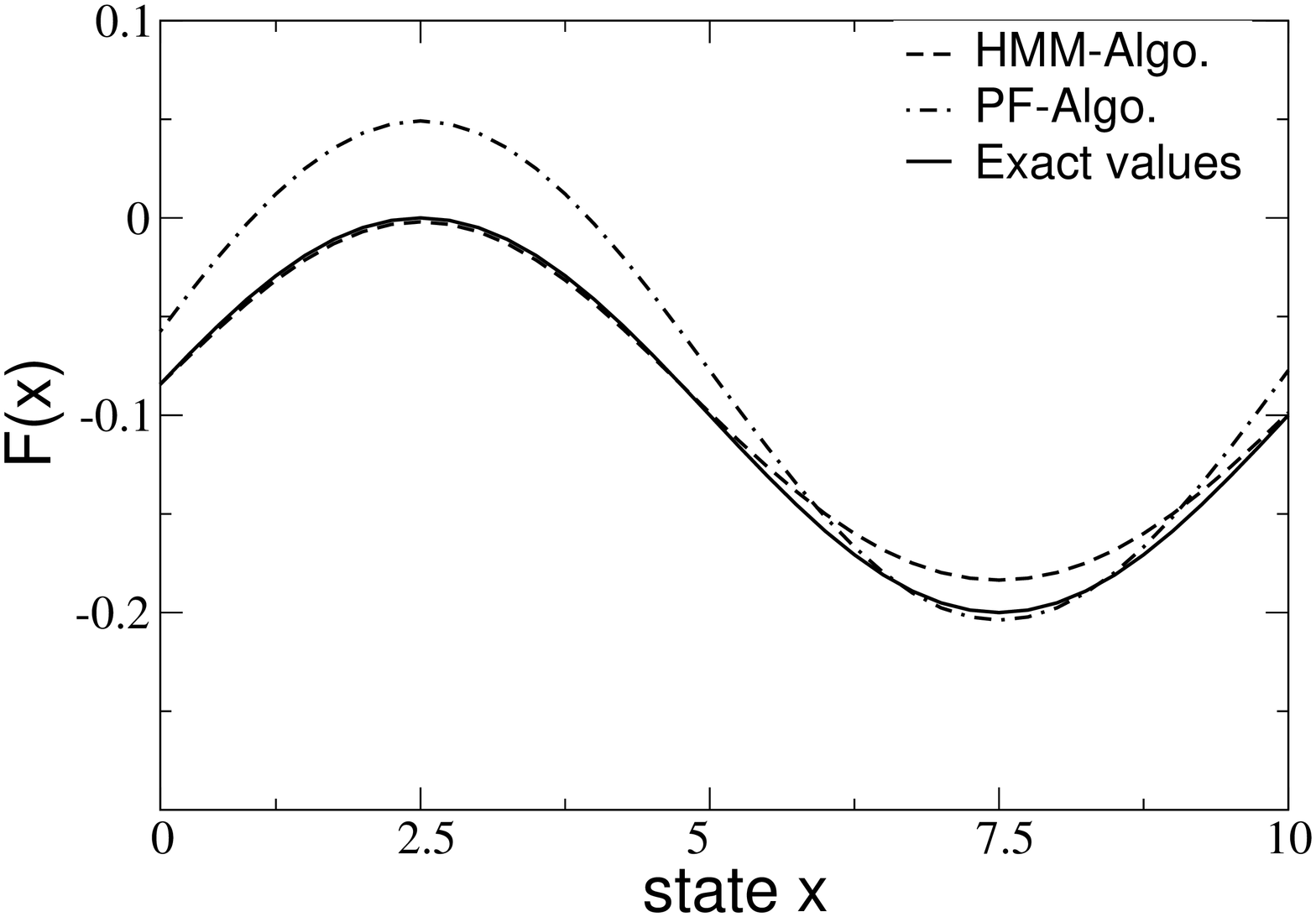,width=8cm,height=8cm,angle=-90}
\psfig{figure=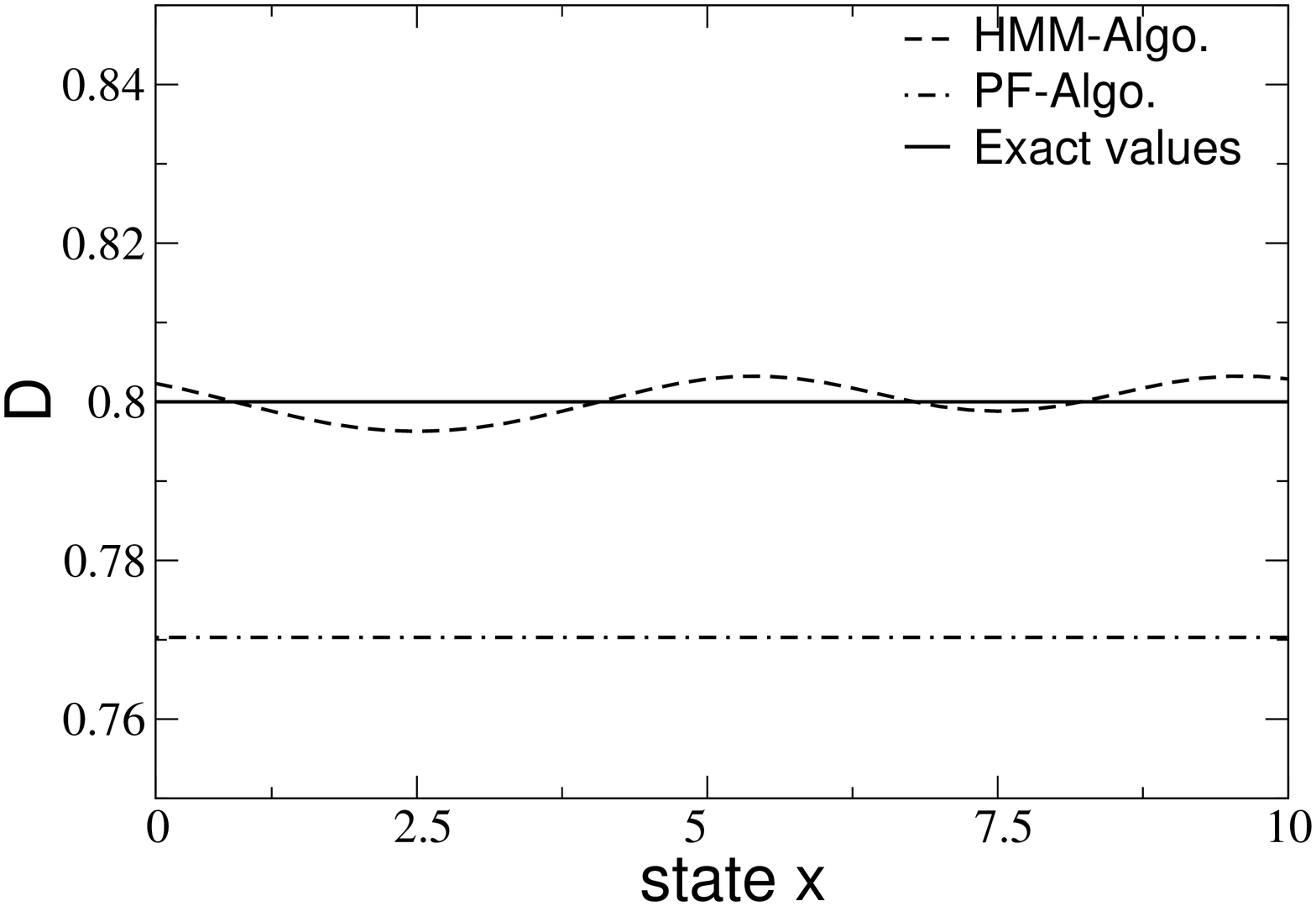,width=8cm,height=8cm,angle=-90}}
\caption{Drift F(x) (left panel) and diffusion coefficient $D$
(right panel) vs. the state. The correct parameters for the drift are
$\theta_0 = -0.1$ and  $\theta_1 =  0.1$, and for the diffusion $D=0.8$ 
which are shown by solid lines. The estimated values given by the modified
Baum-Welch algorithm are shown by dashed lines and the estimated values
given by particle filter are represented by dot-dashed lines.}
\end{figure}
\noindent

Note in Fig. 3 that using the modified Baum-Welch algorithm, the estimation of 
the drift function is better in the interval $x  \leq 5 $, whereas, it is 
better in the domain $x > 5 $ for the particle filter. This is the inverse situation 
if we choose another initial condition.

\section{Conclusion}
In this paper, we have proposed a modified Baum-Welch algorithm based on a
parametrization of the transition matrix associated  with HMMs. This algorithm 
has been compared to particle filters with the aim  to reestimate the 
parameters of hidden diffusion processes 
in periodic potentials and, more precisely, 
to estimate the drift coefficient and the diffusion constant of periodic
stochastic systems.\\
Our simulations show the following results: The particle filter algorithm,
where the number of samples and the length of observation are chosen 
to be large $(N_p = 1000,\, T=1000)$, converges
quantitatively to the correct values of the drift and diffusion
coefficients. The great advantage of the particle filter algorithm is 
its enormous flexibility. It can be applied to practically 
all nonlinear and/or non-Gaussian high-dimensional state space models
within a statistical framework. This algorithm, however,
is stochastic in nature (based on Monte Carlo)
and, it requires a relatively large number of samples
to ensure a fair maximum likelihood estimate of the current state.
In contrast, the modified Baum-Welch algorithm is deterministic and
the transition probabilities between the hidden states
are constrained by the parametrization. 
The modified Baum-Welch algorithm converges
to the correct results within 20-30 iterations of the reestimation procedure.
Thus, the basic idea of this paper works well and the performance in large 
$N$ (continuum limit) can also be evaluated also for more complicated situations.\\

\ack  The authors thank A. L\"oser, H. Kantz and  J. Timmer for helpful discussions.
   This work was partially supported by the DFG-Schwerpunktprogramm 1114
 ``Mathematical methods for time series analysis and digital image processing''.


\end{document}